\documentclass{kluwer}

\usepackage[dvips]{graphics,color,graphicx}

\begin{document}

  \begin{article}

    \begin{opening}

      \title{Solar Wind Forecasting with Coronal Holes}
      \author{\surname{S. Robbins$^1$}}
      \author{\surname{C. J. Henney$^2$}}
      \author{\surname{J. W. Harvey$^2$}}
      \institute{$^1$APS Department, University of Colorado, 
	Boulder, CO 80309, USA \\
	$^2$National Solar Observatory, Tucson, Arizona, 85719, USA}

      \begin{abstract}
	An empirical model for forecasting solar wind speed related geomagnetic events is 
	presented here. The model is based on the estimated location and size of solar 
	coronal holes. This method differs from models that are based on photospheric 
	magnetograms (e.g., Wang-Sheeley model) to estimate the open field line configuration. 
	Rather than requiring the use of a full magnetic synoptic map, the method presented here 
	can be used to forecast solar wind velocities and magnetic polarity from a single coronal 
	hole image, along with a single magnetic full-disk image. The coronal hole parameters 
	used in this study are estimated with Kitt Peak Vacuum Telescope 
	He~{\footnotesize I} 1083~nm spectrograms 
	and photospheric magnetograms. Solar wind and coronal hole data for the period between 
	May 1992 and September 2003 are investigated. The new model is found to be accurate to 
	within $10\%$ of observed solar wind measurements for its best one-month periods, and
         it has a linear correlation coefficient of $\sim$0.38 for the full 11 years studied.
	Using a single estimated coronal hole map, the model can forecast the Earth directed
	solar wind velocity up to 8.5 days in advance. In addition, this method can be used
	with any source of coronal hole area and location data.
      \end{abstract}

   \end{opening}
    
    \section{Introduction}
    Prediction of space weather near the Earth is a major goal of solar
    research. An important aspect of attaining this goal is to accurately
    describe the solar drivers of space weather. The drivers are the
    solar wind and the various phenomena that shape and modulate that wind.
    Among the earliest findings from space observations of the solar wind
    was that it consisted of recurrent low-speed, dense streams and
    high-speed tenuous streams, and that the latter were strongly associated
    with increased geomagnetic activity \cite{Synder63}.
    Many suggestions were made that the high-speed solar wind streams might be
    associated with regions on the Sun having magnetic fields open to
    interplanetary space.
    
    When coronal holes were found to be regions likely
    to have open magnetic fields \cite{Alt72},
    it was not long until \inlinecite{Krieg73}, and several
    other investigators, demonstrated a link between open-field
    coronal holes, high-speed solar wind streams, and enhanced
    geomagnetic activity. In an effort to strengthen this linkage,
    \inlinecite{Sheel76} constructed time-stacked
    diagrams of coronal holes, solar wind speed and
    geomagnetic activity in one-rotation-long rows. The diagrams covering
    the years 1973-1975 strongly supported the linkage and the authors suggested
    that observations of coronal holes could be used to predict the arrival
    of high-speed streams and their associated geomagnetic activity a week
    in advance.

    \begin{figure}
      \includegraphics[width=2.2in]{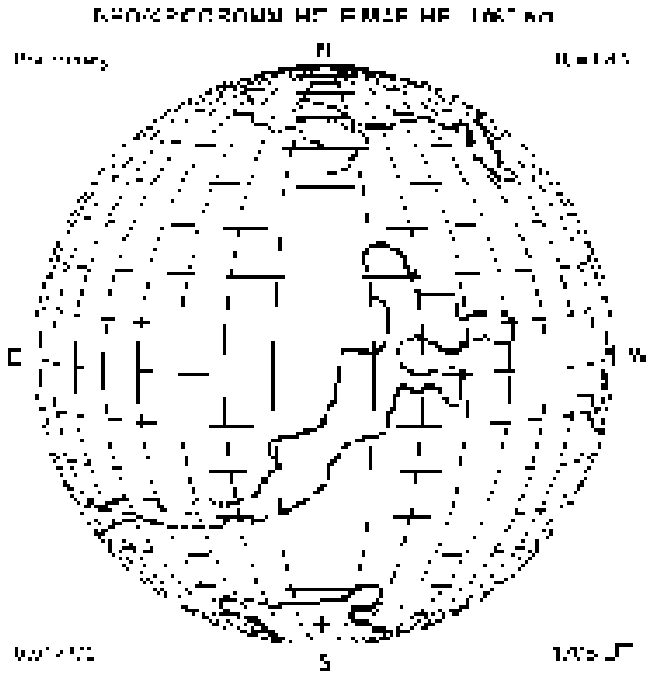}
      \hspace{1pc}
      \includegraphics[width=2.2in]{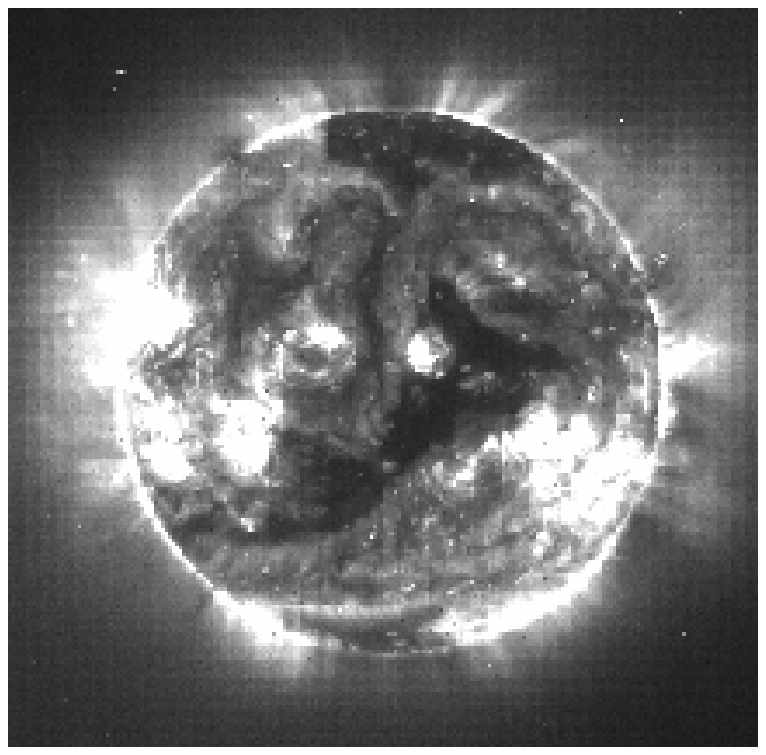}
      \caption{A sample NSO/KPVT computer-assisted hand-drawn 
	coronal hole image (left) and a EIT 19.5 nm Fe XII emission line 
	image (left) for July 14, 2003 at approximately 17 UT.
	Note that the coronal hole regions appear dark in the EIT image.}
    \end{figure}
    
    Coronal holes are best seen against the solar disk as low-intensity
    regions in space observations of material at coronal temperatures. This
    can also be done from the ground using radio observations. \inlinecite{Harvey75}
    found that coronal holes could be seen faintly in ground-based images
    made with helium lines such as 587.6 and 1083.0 nm because the strength
    of these lines is partly controlled by the intensity of overlying
    coronal radiation (see, e.g., \opencite{And97}). A program of regular 1083 nm observations 
    has been conducted by the National Solar Observatory (NSO) Kitt Peak Vacuum 
    Telescope (KPVT) starting in 1974. 
    Among the derived products are estimates of the locations and magnetic polarity 
    of coronal holes. An example coronal hole estimate image derived from a KPVT
    observation is shown in Figure~1, along with a 19.5 nm Fe XII emission 
    line image measured by the Extreme Ultraviolet Imaging Telescope (EIT)
    for comparison. \inlinecite{KHar02} describe how coronal holes are identified 
    using KPVT He~{\footnotesize I} 1083~nm observations. 
    
    Predictions of solar wind speed at Earth are regularly made by several
    groups based on solar potential field extrapolations (e.g., \\ 
    http://solar.sec.noaa.gov/ws/, http://www.lmsal.com/forecast/, \\
    http://bdm.iszf.irk.ru/Vel.html, and http://gse.gi.alaska.edu)
    and interplanetary scintillation \cite{Hew1964}
    observations (e.g., \\
    http://cassfos02.ucsd.edu/solar/forecast/index.html, and \\
    http://stesun5.stelab.nagoya-u.ac.jp/forecast/). 
    The former set of forecasts are based on extrapolation of photospheric
    longitudinal magnetic field measurements using a potential field
    assumption to locate open field lines. Using one of these models,
    a modified \citeauthor{Wang90} (\citeyear{Wang90,Wang92}) flux-transport model,
    \inlinecite{Arg00} studied a three-year period centered about the
    May 1996 solar minimum. They compared predicted solar wind speed and
    magnetic polarity with observations near Earth. Their three-year
    sample period had an overall correlation of $\sim0.4$ with observed
    solar wind velocities and an average fractional deviation, $\xi$, of $0.15$, where
    $\xi = \left\langle {{({\rm{prediction - observed})}}/{{\rm{observed}}}}
    \right\rangle$. When excluding a 6-month period with
    large data gaps, they correctly forecast the solar wind to within
    $10-15\%$. Interplanetary magnetic field (IMF) polarity was correctly
    forecast $\sim75\%$ of the time.

    In this paper, we address the suggestion of \inlinecite{Sheel76}
    that observations of coronal hole regions can be used to predict the
    solar wind speed at Earth as much as a week in advance. In addition,
    the model presented here is based on observations that find moderate
    and high-speed solar wind streams are associated with small and large
    near-equatorial coronal holes, respectively \cite{Nolte1976}. Here we correlate 
    the coronal hole percent area coverage of sectoral regions of the observed 
    solar surface with solar wind measurements to derive a simple empirical
    model (discussed in Sections 2 through 5). As a measure of the merit 
    of this model for solar wind forecasting, we compare
    predictions with observations and contrast this technique with the ones
    based on magnetic field extrapolations (Sections 5 and 6).

    \section{Model Input Data} \label{data}
    The coronal hole data used here are based on KPVT observations from May 28, 1992 
    through September 25, 2003 (i.e. the last half of cycle 22 and the first half of 
    cycle 23). The coronal hole locations and area estimates are from 
    computer-assisted, hand-drawn maps (see Figure~1) based upon the 
    KPVT He~{\footnotesize I} 1083~nm images and photospheric magnetograms \cite{KHar02}.
    For this investigation, the estimated coronal hole boundaries 
    were mapped into sine-latitude and longitude to create heliographic images. 
    The coronal hole region image pixels are set to a value of~1, 
    whereas the background is defined as 0. For the time period analyzed here, the KPVT 
    coronal hole maps have a $69\%$ daily coverage.
    
    The solar wind speed data utilized here was obtained from the OMNIWeb 
    website (http://nssdc.gsfc.nasa.gov/omniweb/) provided by the National Space Science 
    Data Center.  Daily averages of the solar wind speed time series were created with the 
    approximate cadence of the KPVT-based coronal hole maps. For the time period analyzed 
    here, the solar wind speed time series has a $92\%$ daily coverage. Data gaps in the time 
    series are interpolated using a cubic spline.

    \section{Solar Wind Correlation Analysis}\label{analysis}
    For comparison with the solar wind speed time series, each heliographic coronal 
    hole image was divided into 23 swaths (i.e. sectoral regions) $14$ degree-wide 
    in longitude overlapped by 7 degrees. The approximately $1$-day-wide longitudinal window 
    was selected to correspond with the temporal cadence of the KPVT observations. These
    sectoral samples are then summed, where each pixel corresponding to a coronal hole
    is equal to 1, to yield a percent coverage of that area by coronal holes. 
    For each coronal hole image there may be no or only a few coronal hole regions 
    observed for that time. For example, swath sectors with no coronal hole regions 
    would yield a hole coverage of zero percent.
    This is repeated for each coronal hole image available in the $11$-year period to form 
    a coronal hole time series for each of the 23 sectoral samples. Each sectoral time series is 
    then interpolated into the time frame of the solar wind velocity data.
    
    The correlation and time lag between the time series were estimated with
    weighted cross-correlations (e.g. \opencite{bev03}). The weighted cross-correlation
    simplifies the analysis by allowing the use of the continuous time series.
    The gap-filled data are given small weights to minimize their contribution while 
    the measured or derived values are given equal and relatively large weight
    values. In addition, following \inlinecite{Arg04}, periods of CME events were 
    estimated using the plasma $\beta$ value (obtained from the OMNIWeb data set) 
    when $\beta  \le  0.1$. For periods estimated to correspond to a coronal mass 
    ejection (CME) event, solar wind speed values were given negligible weight values. 

    \begin{figure}
      \includegraphics[width=4.8in]{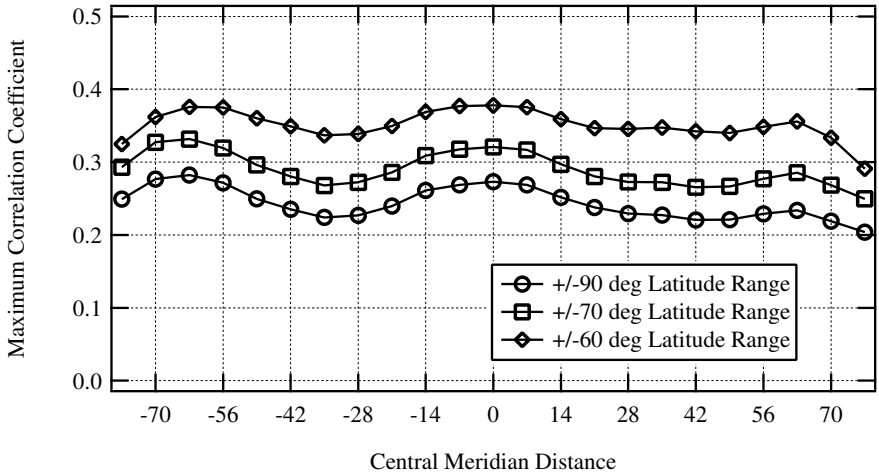}
      \caption{Longitude Cross-Correlations:  $14$ degree-wide swaths of the solar disk 
	were tiled across the solar surface with $7$ degrees overlap between successive swaths
	for 3~latitude ranges. The maximum cross-correlations within a lag 
	window of $\pm 0.5$~days (corresponding to a day offset described 
	by Equation (\ref{eq_lagperiod})) of the resulting data set with the solar wind velocity 
	data are shown here.  Negative and positive longitude values correspond to 
	eastward (E) and westward (W) respectively. Note the two relatively good forecast 
	windows centered around $63$ degrees east and the central meridian. The
	correlations continue to improve with model time series using decreasing latitude 
	ranges, peaking with the latitude range of $\pm 60$ degrees. The trend reverses 
	for latitude ranges narrower than $\pm 60$ degrees.}
      \label{fig_longitude}
    \end{figure}

    \subsection{Longitudinal Forecasting Windows}\label{longitude}

    Twenty-three longitude swaths were examined, ranging from $77$ degrees east 
    to $77$ degrees west of the central meridian.  Each of these was cross-correlated with 
    the solar wind speed time series in the manner described above. Figure~{\ref{fig_longitude}} 
    exhibits the maximum cross-correlation coefficient within a lag window of $\pm 0.5$~days
    for each longitude swath investigated with the
    latitude ranges $\pm 90$ degrees, $\pm 70$ degrees, and $\pm 60$
    degrees with equal latitude weights. Note that besides the central meridian peak,
    the correlation also peaks towards the east and west limbs. This is a result of better 
    coronal hole detection from the KPVT  He~{\footnotesize I} 1083~nm spectroheliograms 
    towards the image limb. In addition to the three latitude bands shown
    in Figure~{\ref{fig_longitude}}, the correlation values were also determined for  
    $\pm 50$ and $\pm 40$ deg cases. These two latitude bands are similar to the $\pm 60$ degree
    case but have lower correlation values eastward of -56 degrees. For the $\pm 60$ deg case,
    two preferred forecasting windows centered at $63$ degrees east and on the central meridian 
    are clearly visible. 

    The estimated swath time series lags corresponding to maximum cross-correlation
    coefficients are found to be linear, and can be expressed as:
 
    \begin{equation}\label{eq_lagperiod}
      d = \left( {3.69 \pm 0.02} \right) - \left( {0.07386 \pm 0.0005} \right)\theta,
    \end{equation}
    
    \noindent where $d$ is the time delay in days and $\theta$ is the center of the 
    longitude swath in degrees as measured from the central meridian (east is negative, 
    west is positive) and the uncertainty values are $1$-$\sigma$. Note that Equation (1)
    is valid within the central meridian distance range of -80 to 80 degrees, but it only has
    physical meaning (days forecast) for the central meridian distance range of -80 to 40
    degrees. From Equation (1), 
    the two preferred forecasting windows centered at $63$ degrees east and 
    on the central meridian (see Figure~{\ref{fig_longitude}}), the time lags of 
    these forecasts correspond to $8.3$ days and $3.7$ days respectively. In other words,
    the delay between detected solar wind variations at Earth and a coronal hole region at 
    the central meridian ($\theta = 0$ degrees) is approximately $3.7$ days. 
    A coronal hole region observed at $63$ degrees east central meridian distance 
    ($\theta = -63$ degrees) is associated to solar wind speed variations at Earth 
    approximately $8.3$ days later. The observed delay is as expected, and is best explained
    as the result of a corotating stream of plasma moving nearly radially outward 
    from the sun (e.g. \opencite{Gos1996}).
    
    \begin{figure}
      \includegraphics[width=4.8in]{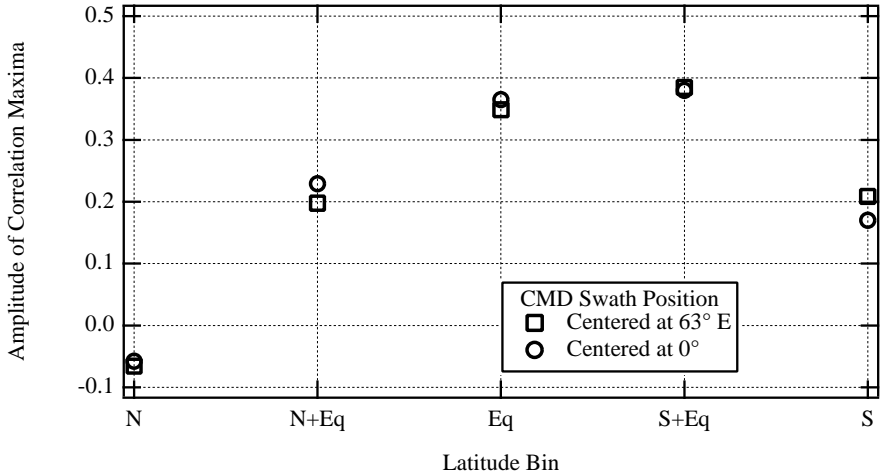}
      \caption{The two longitude forecasts' cross-correlations 
	with the solar disk divided into five latitude regions described in \S \ref{latitude}. 
	See text for discussion.}
      \label{fig_latitude_used}
    \end{figure}

    \subsection{Latitude Weighting Analysis}\label{latitude}
    In addition to the longitudinal correlation analysis above, the heliographic
    coronal hole images were divided into three latitude bins: Northern ($90$ degrees north to 
    $30$ degrees north), Equatorial ($30$ degrees north to $30$ degrees south), and Southern 
    ($30$ degrees north to $90$ degrees south) regions.  
    Combinations of Northern with Equatorial and Southern with Equatorial were used for 
    a total of five bins.  The cross-correlation coefficient values of each latitude bin were 
    calculated for the two longitudinal windows discussed in \S \ref{longitude}.

    The amplitude of the correlation coefficient maxima shown in 
    Figure~\ref{fig_latitude_used} reveals a 
    bias towards the southern hemisphere. This bias in the cross-correlation 
    between the hemispheres
    is most likely due to the significantly greater number of coronal holes detected in the 
    southern hemisphere for the period investigated (e.g. \opencite{Henn05}). 
    Assuming that the hemispheric asymmetry is a result of the limited
    distribution sample of coronal holes, and following the analysis discussed in
    the previous section, the latitude range of $\pm$60 degrees was used for the 
    forecasting analysis done below. The maximum weighted 
    cross-correlations for the East window was $0.376$
    and the central meridian had a correlation of $0.378$ (see Figure \ref{fig_longitude}).
    
    A model solar wind time series, $V_{\rm{mod}}$, is created by first 
    determining the area percentage of the 14-degree wide sectors that is a coronal 
    hole, $I_{\rm{s}}$. These coronal hole percentage values for each longitudinal 
    swath are rescaled to agree with the observed solar wind speed, $V_{\rm{obs}}$, using 
    the linear scaling coefficients $\alpha$ and $\gamma$, where $V_{{\rm{mod}}} 
    = \alpha  + \gamma I_{{\rm{s}}}$. The linear scaling coefficients were determined by 
    minimizing the mean of the absolute average fractional deviation, $\xi$, of the model from 
    the observed values, where
    
    \begin{equation}\label{eq_mafd}       
      \xi = \left\langle 
	  {{{\left( V_{\rm{obs}} - V_{\rm{mod}} \right) / V_{\rm{obs}} }}}
	  \right\rangle .
    \end{equation}
    
    \noindent Using the above criteria and a latitude range of $\pm 60$ degrees, the 
    average linear scaling values for each longitude window are found to 
    be: $\alpha = 330~{\rm km/s}$ and $\gamma = 930~{\rm km/s}$. With these scaling factors and
    the best weighting as discussed above, the resulting $\xi$ for the entire
    $11$-year data set for the central meridian swath is $\sim 16\%$ with a standard 
    deviation of $\pm 20\%$.

    \begin{figure}
      \includegraphics[width=4.8in]{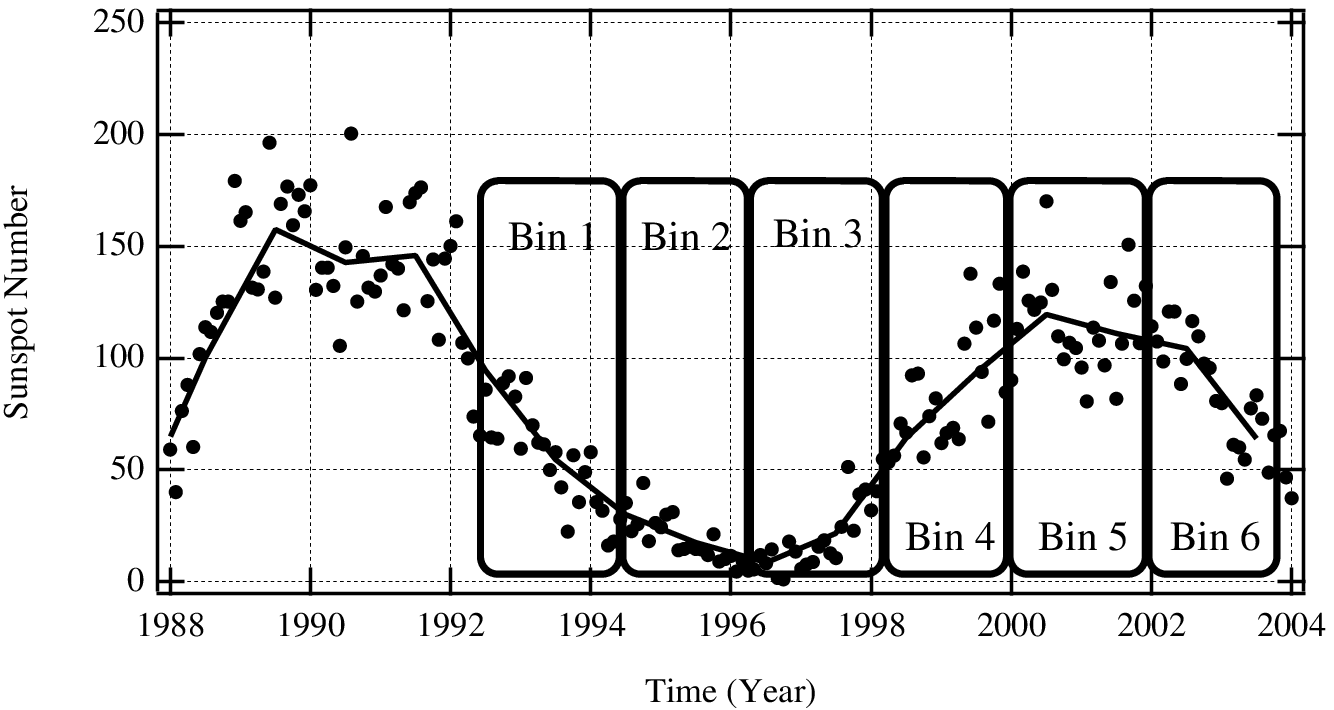} \\
      \includegraphics[width=4.8in]{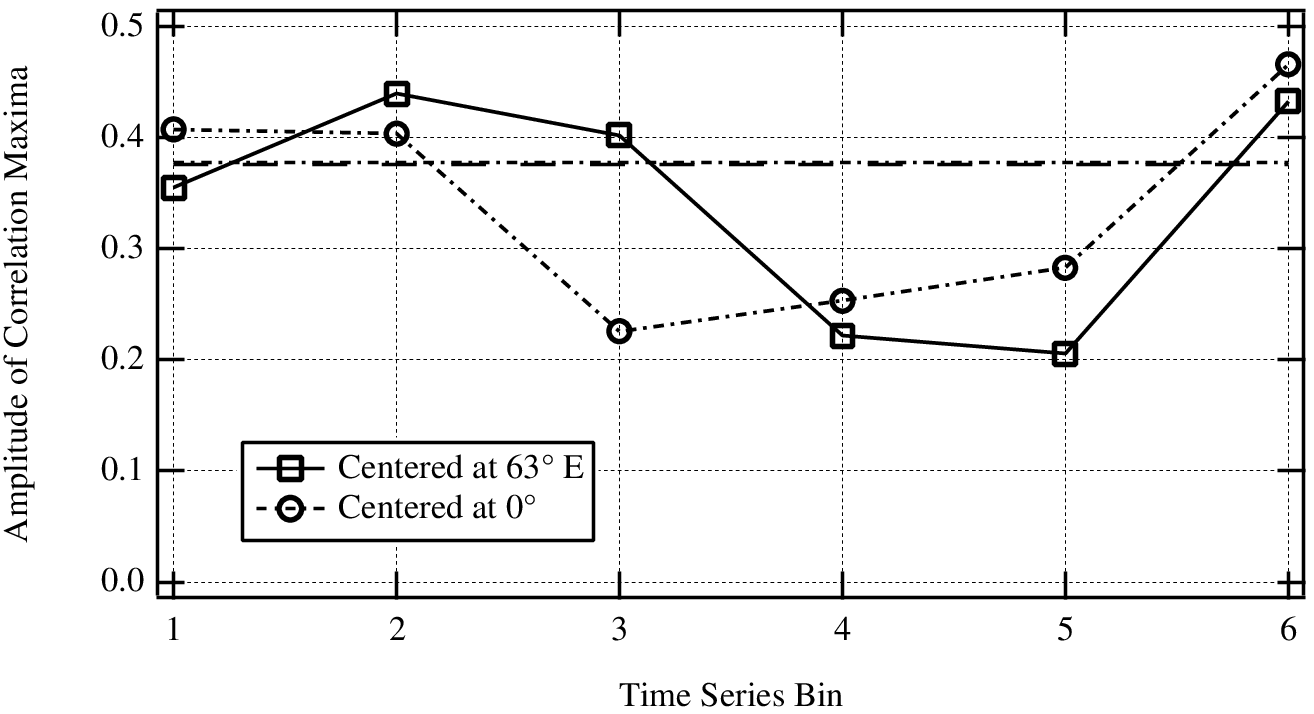}
      \caption{Monthly (solid dots) and yearly (line) sunspot number averages are shown 
	in the upper panel along with six time bins into which the 11-year data set was divided. 
	The six bins are indicated by the rounded rectangles and each represents a duration of 
	approximately $690$ days. 
	Date ranges for each period are listed in Table \ref{table_timeseries}. Shown in the lower 
	panel are the cross-correlation 
	values of the coronal hole data with the solar wind for the 6 intervals illustrated in 
	the upper figure. The correlation is found to be best during the declining phase of the
	sunspot cycle and worst just after solar minimum.}
	\label{fig_cycledependence}
    \end{figure}

    \section{Magnetic Activity Cycle Dependence} \label{sunspotcycle}
    
    \inlinecite{Sheel81} reported a dependence between the coronal hole and solar wind
    correlation and the sunspot cycle. This was 
    quantitatively explored for the time period spanned by the coronal hole data set - the last 
    half of cycle 22 and the first half of cycle 23.  The full time series was divided 
    into six time series of approximately 690 days, illustrated in the
    upper graph in Figure \ref{fig_cycledependence} with sunspot number time series
    (sunspot count data used here was obtained from the NGDC website (http://www.ngdc.noaa.gov/stp/),
    maintained by the National Geophysical Data Center). The date range and percent coverage 
    of the six periods are outlined in Table \ref{table_timeseries}.

    \begin{table}
      \caption{Time series subdivisions used in the cross-correlation analysis relative
	to the sunspot number time series shown in Figure~\ref{fig_cycledependence}.}
      \begin{tabular}{clcc}
	\hline
	Period Bin &Date Range & Interval (days) & Completeness \\
	\hline
	1 &May 28, 1992 - Apr 17, 1994 & 690 & 66.1\% \\
	2 &Apr 18, 1994 - Mar 7, 1996 & 690 & 68.4\% \\
	3 &Mar 8, 1996 - Jan 26, 1998 & 690 & 72.2\% \\
	4 &Jan 27, 1998 - Dec 17, 1999 & 690 & 73.8\% \\
	5 &Dec 18, 1999 - Nov 6, 2001 & 690 & 64.6\% \\
	6 &Nov 7, 2001 - Sep 25, 2003 & 688 & 66.3\% \\
	\hline
	1 - 6 &May 27, 1992 - Sep 25, 2003 & 4138 & 68.6\% \\
	\hline
      \end{tabular}
      \label{table_timeseries}
    \end{table}

    Figure \ref{fig_cycledependence} highlights a strong dependence of the time series 
    correlation values on the phase of the sunspot cycle, similar to the qualitative results of 
    \inlinecite{Sheel81}. The correlation between the coronal hole and solar wind time series
    is best during the declining phase of the cycle, and it is worst during solar minimum and 
    the beginning ascending phase of the cycle. Some of the lack of correlation can be attributed
    to the observed latitudinal distribution of coronal holes with respect to the solar cycle.
    During the minimum phase of the solar cycle, fewer low-latitude coronal holes are observed
    which means few fast streams are observed by spacecraft in the ecliptic plane (e.g. \opencite{Woch1997}; 
    \opencite{Koj2001}; \opencite{Koj2004}).
    
    However, the source of the poor correlation during solar minimum is also likely a result of
    a noted difficulty in the determination of coronal hole regions using He~{\footnotesize I} 1083~nm 
    spectroheliograms during periods of low magnetic activity.
    Both longitude windows vary over approximately the same range throughout the cycle, and even though
    during the worst forecasting period the correlation is far below the full time series' average,
    it is still well above statistical significance. Correlation significance,
    the probability that two uncorrelated random sets of variables with a given number of observations 
    would give similar correlation values, was calculated following Appendix C in \inlinecite{bev03}.  
    The correlation is considered highly significant and nominally significant
    if the probability of chance occurrence is less than $1\%$ and $5\%$, respectively \cite{Taylor97}.
    
    The time delays between coronal hole observation and the effects seen in the solar wind
    velocity are slightly longer during solar minimum and slightly shorter during solar maximum 
    than the averages quoted in \S \ref{longitude}.  The lags range over $8.38 \pm 0.34$ and
    $3.82 \pm 0.58$ days.

    \section{IMF Polarity Forecast} \label{polarity}
    
    Besides area and position information, the \inlinecite{KHar02} coronal hole 
    boundary images include magnetic polarity. So, in addition to forecasting the 
    solar wind velocity, the KPVT estimated coronal hole maps can be used to predict the 
    interplanetary magnetic field (IMF) polarity at Earth.  The longitudinal swath centered at
    the central meridian, the $3.7$-day forecast, was used in the following analysis.
    The heliographic coronal hole maps are scaled so that each pixel with a positive polarity 
    hole has a value of $+1$ and each pixel with a negative polarity hole has a value 
    of $-1$, whereas non-coronal hole regions are set to $0$. The average value of all the 
    pixels in the $14$ degree wide swath was taken, with a range between $\pm60$ degrees in
    latitude. The averaging of the coronal hole polarity, albeit simple, is treated here a
    as baseline for the polarity forecasting when using only coronal hole regions.
    
    Excluding only the days that did not have both velocity and coronal hole data, leaving $63.0\%$ 
    of the comparison period, the IMF polarity was correctly forecast $57.9\%$ of the time.  When
    excluding an additional $17.9\%$ of the days where the average model magnetic polarity was 
    within $0.1\%$ of $0$, the IMF polarity is correctly forecast for $63.5\%$ of the time. This 
    is approximately $10\%$ lower than the values reported by \inlinecite{Arg00}; however, we 
    expect improvement with future models. Though the current model has inherent 
    inaccuracies as a result of the oversimplification of the magnetic field structure associated
    with coronal holes and the resulting solar wind, the use of higher signal-to-noise ratio 
    magnetograms is expected to improve the polarity forecast. In addition, this model is partly
    based on the assumption that the solar wind velocity is related linearly with the size of the
    coronal hole. However, it has been shown that there is a critical scale size for which the 
    wind velocity is independent of coronal hole size \cite{Koj2004}. In future models, we plan 
    to include coronal hole size and additional topology-related parameters to potentially improve 
    forecasts.
    
    \begin{figure}
      \includegraphics[width=4.8in]{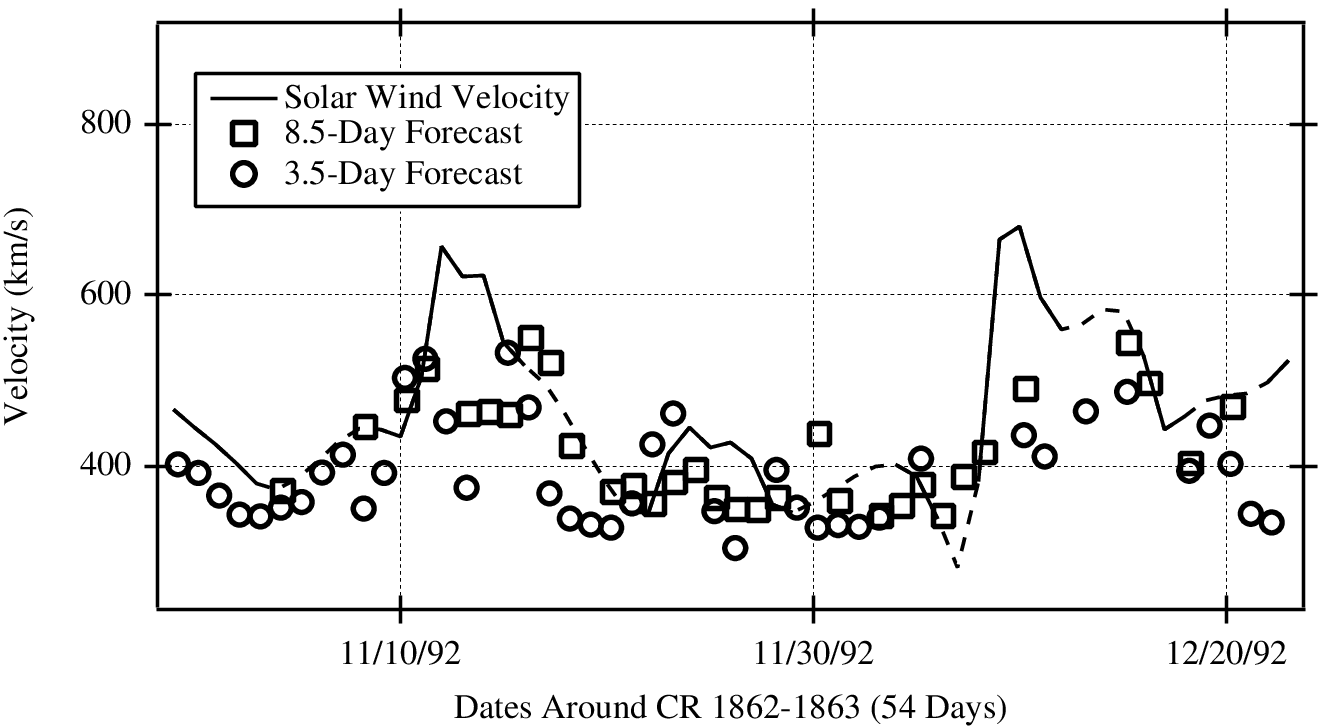} \\
      \includegraphics[width=4.8in]{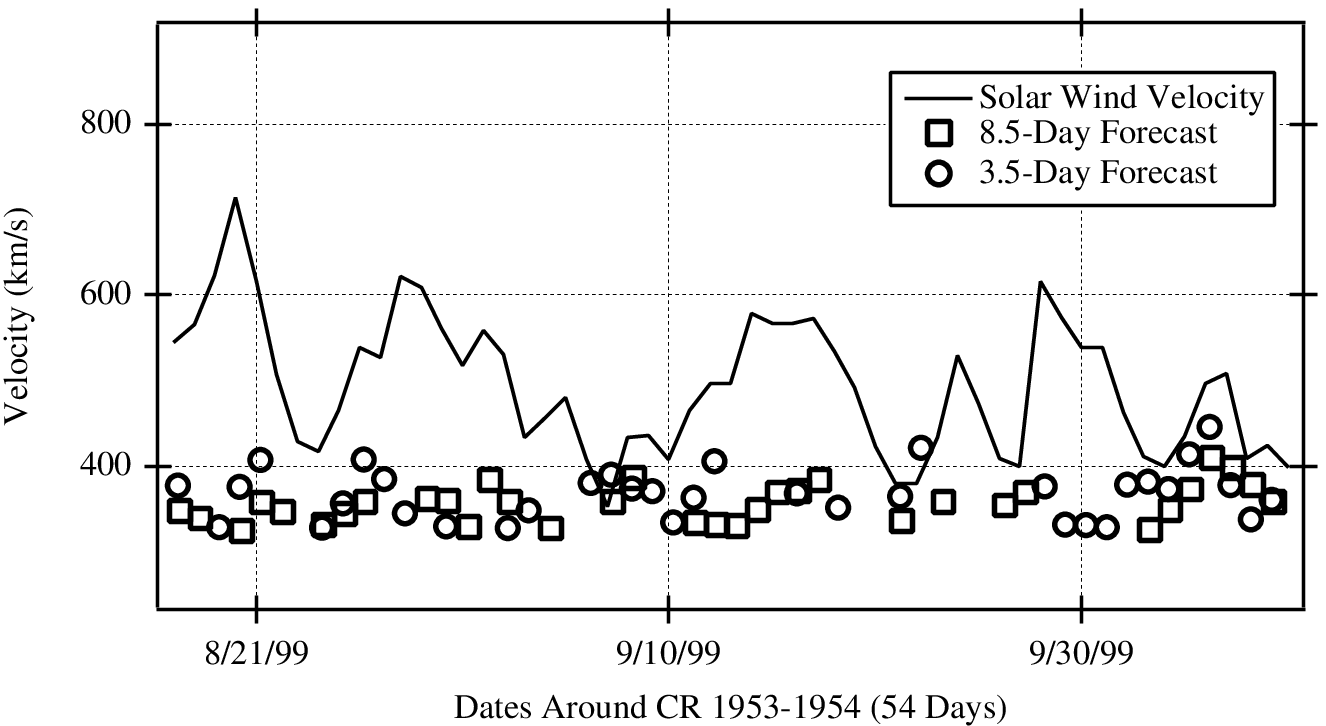}
      \caption{A good forecast (top) comparison relative to a representative poor forecast (bottom)
	comparison between the model estimates (symbols) and the measured solar wind
	speed (solid line).
	In the top forecast, dotted lines are interpolated solar wind speed values.}
      \label{fig_forecastexamples}
    \end{figure}

    \section{Forecast Comparison and Discussion} \label{examples}

    Figure \ref{fig_forecastexamples} illustrates two sample forecast periods that cover two 
    Carrington rotations (CR) each: CR 1862 and 1863 (top), and CR 1955 and 1954 (bottom).
    To objectively find periods of good and bad forecasts, both weighted cross-correlations for
    $90$-day periods as well as absolute average fractional deviations, see Equation
    (\ref{eq_mafd}), were performed. For each time series, only valid data (non-gap-filled 
    data values) are shown in the figures.  Ranges quoted
    in this section are from the two different forecast windows (discussed in \S \ref{analysis})
    centered at $63$ degrees east and $0$ degrees relative to the central meridian.
    
    The top forecast comparison shown in Figure \ref{fig_forecastexamples} is a sample period 
    when the model time series is well-correlated with
    the solar wind speed data.  For the two forecast windows, the weighted correlation coefficients
    range from $0.626$ to $0.698$, and the absolute average fractional deviation, $\xi$, is found to be 
    $0.098$.  In comparison, the
    best one-month period studied by \inlinecite{Arg00} has an unweighted correlation of $0.813$
    and an absolute average fractional deviation of $0.159$.
    The bottom forecast comparison shown in Figure \ref{fig_forecastexamples} illustrates an
    example of a poor cross-correlation between the two longitude windows and the solar
    wind velocity, ranging from $0.113$ to $0.179$.  This is below statistical significance
    for the forecast comparison period of $54$ days (shown in Figure \ref{fig_forecastexamples}).
    The absolute average fractional deviation ranges from $0.207$ to $0.245$ for this period.
    
    For the three-year period studied by \inlinecite{Arg00}, they found, using their best
    forecast method, a correlation of $0.389$, whereas the entire $11$ years studied here has 
    a weighted correlation range of $0.376$ to $0.378$ depending on longitudinal window used.    
    In addition, \inlinecite{Arg00} reported an absolute average fractional deviation, $\xi$, for 
    the three years studied of $0.15$.
    The best absolute average fractional deviations for a one-month period presented in their 
    paper was $0.096$. The weighted $\xi$ for the best one-month
    periods studied here ranged between $0.073$ and $0.076$ for the two forecast windows.
    Months with less than $10$ days for which the data sets
    overlapped after weighting were excluded.  For the approximate $137$ months studied
    in this paper, the
    mean $\xi$ ranged between $0.167$ and $0.176$ with a standard deviation of about $0.051$.
    
    Near term plans include applying the forecasting model presented here to SOLIS-VSM (Vector 
    Spectromagnetograph) estimated coronal hole images derived from daily full-disk photospheric 
    magnetograms and He~{\footnotesize I} 1083~nm spectroheliograms using an automated 
    coronal hole detection algorithm developed by \inlinecite{Henn05}.

    \section{Conclusion}
    
    The empirical solar wind forecasting model presented here is based on the location and size 
    of solar coronal holes. From a single coronal hole area image, along with a single 
    magnetogram, this method can be used to 
    forecast solar wind velocities and polarity up to $8.5$ days in advance with a mean wind 
    speed deviation as low as $9.6\%$ for a given one-month period. The model is found to be 
    accurate to within $10\%$ of observed solar wind measurements for its best one-month periods.    
    Possible improvements include adding the estimated quality of the coronal hole
    area determination, along with weighting by the size and topology of each coronal hole, used 
    as input to the model.

    \section{Acknowledgments}

    The coronal hole data used here was compiled by K. Harvey and F. Recely 
    using NSO KPVT observations under a grant from the 
    National Science Foundation (NSF). NSO Kitt Peak data used here are produced 
    cooperatively  by NSF/AURA, NASA/GSFC, and NOAA/SEL. The EIT data is courtesy 
    of the SOHO/EIT consortium. SOHO is a project on international cooperation between 
    ESA and NASA. Solar wind and plasma $\beta$ data used here is available online at 
    http://nssdc.gsfc.nasa.gov/omniweb/.  Sunspot data used here is available online at
    http://www.ngdc.noaa.gov/stp/.  This work is carried 
    out through the NSO Research Experiences for Undergraduate (REU) site program, 
    which is co-funded by the Department of Defense in partnership with the NSF-REU Program.  
    This research was supported in part by the Office of Naval Research Grant 
    N00014-91-J-1040. The NSO is operated by AURA, Inc. under a cooperative agreement 
    with the NSF.
    


  \end{article}
\end{document}